\documentclass[a4paper,fleqn,usenatbib]{mnras}

\usepackage{newtxtext,newtxmath}

\usepackage[T1]{fontenc}
\usepackage{ae,aecompl}

\usepackage{graphicx}	
\usepackage{amsmath}	
\usepackage{amssymb}	
\usepackage{enumerate}

\title[The nature of GeMS distortions]{Unveiling the nature of the Gemini multiconjugate adaptive optics
system distortions}

\author[M. Patti \& G. Fiorentino]{
M. Patti$^{1}$\thanks{E-mail: mauro.patti@inaf.it}
 and G. Fiorentino$^{1}$
\\
$^{1}$INAF - Osservatorio di Astrofisica e Scienza dello Spazio di Bologna, Italy\\
}

\date{Accepted 2019 February 25. Received 2019 February 21; in original form 2018 December 28}

\pubyear{2018}

\begin{document}
\label{firstpage}
\pagerange{\pageref{firstpage}--\pageref{lastpage}}
\maketitle

\begin{abstract}
Astrometry was not a science case of the Gemini Multiconjugate adaptive optics System (GeMS) at its design stage. 
However, since GeMS has been in regular science operation with the Gemini South Adaptive Optics Imager (GSAOI), their astrometric performances have been deeply analysed. 
The non-linear component of the distortion map model shows a characteristic pattern which is similarly repeated in each detector of GSAOI.
The nature of this pattern was unknown and subjected to different hypotheses. This paper describes the origin of the GeMS distortion pattern as well as its multi-epoch variation. At the end, it is showed a comparison with the current design of the Multiconjugate Adaptive Optics RelaY (MAORY) of the Extremely Large Telescope (ELT). 
\end{abstract}




\section{Introduction}
Detailed and systematic astrometric studies of stellar kinematics in our Galaxy and in nearby stellar systems can reveal the imprints of their formation and dynamical evolution. Since the diffraction limit decreases with the telescope diameter, it is obvious that the so called "extremely large telescopes" (ELTs) are expected to have superb astrometric performance that will not be overcame for the next decades \citep[e.g.][]{fiorentino2017}.
Adaptive optics (AO) is the sophisticated technique that will allow these giants to reach their diffraction limit from the ground. The main limits of classical AO observations are:
\begin{itemize}
\item the presence of a very bright star (typically  R $\le$ 11 mag) close to the scientific target (Single Conjugate Adaptive Optics, SCAO); 
\item the small Field of View (FoV), i.e. $\sim$20 arcsec$^2$. 
\end{itemize}
The Multi Conjugate AO (MCAO)~\citep{beckers88,beckers89a} strongly mitigates these problems by using two or more fainter (R$\le$ 13.5 mag) stars to perform the atmospheric turbulence correction, thus increasing the size of the corrected FoV up to 1 arcmin$^2$.\\
Given an uncrowded stellar field, observed at a fixed signal-to-noise ratio (SNR), the centroiding error is proportional to the full width at half maximum (FWHM) of the star. Conversely, fixing the FWHM, the centroiding error decreases as the SNR increases. When compared with seeing-limited observations, an MCAO assisted camera can detect fainter sources at their diffraction limit (minimum FWHM) therefore it is very attractive for astrometric studies \citep{massari2016astrometry}.
A large number of high-quality and uniform Point Spread Functions (PSFs) reduces significantly not only the centroiding errors but also the fitting errors of high order distortion models, thus posing the basis for high precision astrometry. Furthermore, a large FoV has also an impact on astrometric accuracy since it increases the probability to find astrometric reference sources close to the scientific target.\\
A notable target for astrometric studies is the Galactic center. A huge amount of observing time has been devoted to this target with several telescopes and AO flavours \citep[e.g.][and references therein]{gillessen17}. In order to improve the level of precision and accuracy of stellar proper motions in the Galactic center, particular attention has been dedicated to remove geometric distortions of the camera and to build a robust on-sky astrometric reference frame, see ~\cite{yelda2010improving} for a detailed discussion. These authors analysed the impact on astrometric precision and accuracy of removing both the static camera distortions (by using HST data) and the dynamical distortions typical of AO observations (by using SiO masers). 
They conclude that a reasonable number of astrometric calibrators within the FoV improves not only accuracy but also the precision possible with AO observations. It is worth to mention that since the  Gaia~\citep{refId0} mission is surveying the sky, a growing number of stars have well measured proper motions and can serve as astrometric calibrators. Interestingly, the number of expected Gaia reference stars over 1 arcmin$^2$ FoV has been recently estimated ~\citep{rodeghiero2018impact} to asses the astrometric requirement of the MCAO Imaging Camera for Deep Observations ~\citep[MICADO][]{davies2018micado}, the first light high resolution camera for ELT. These authors found about 10 stars when G$\sim$18 mag and SNR$\sim$500. This number is sufficient to model geometrical distortions up to the third order, thus assuring us a solid astrometric reference frame.\\
GeMS~\citep{rigaut2013gemini} is the MCAO system of the Gemini south observatory and currently the most powerful AO module operating on sky. Since it has been in operation, astrometry programs have become a growing fraction of science cases and, even if GeMS was not designed to fulfill astrometric requirements, it is able to reach a high astrometric precision of about 0.2 mas for bright stars on single-epoch observations~\citep{neichel2014astrometric}.\\
To deliver science data, GeMS works with GSAOI in the wavelength interval of 0.9 - 2.4 $\mu$m over a field of view of 85''$\times$85''. The instrument focal plane is covered by an array of four detectors with a pixel scale of 0.02''~\citep{carrasco2012results}.\\
In general, MCAO-assisted cameras are affected by several distortion effects that change the stars coordinates on the detectors~\citep{trippe2010high}. Several authors have analysed GeMS data in combination with space telescope data in order to establish the astrometric performance reached with GeMS and to constrain its geometric distortions ~\citep{massari2016astrometry,dalessandro2016gems,bernard2018optimal}. These studies agree that the GeMS-GSAOI detector shows a distortion structure that is repeated in the four chips separately.
A similar, in shape and magnitude, distortion pattern has been found by using a completely independent approach: a calibration mask was installed at the GeMS entrance focal plane. This mask offers about 1600 pinhole sources to deliver an astrometric distortion correction~\citep{riechert2018gems}. None of these studies try to explain the origin of this geometric distortion.\\ 
In our  paper, we analyse from an opto-mechanical point of view the GeMS geometrical distortion structure. After a brief description of the optical design, Section~\ref{sec:1} highlights the method used to analyse the optical distortions. In Section~\ref{sec:2} we discuss the dependence of geometrical distortions on the observing strategy, single versus multi-epoch observations. Moving to the future, Section~\ref{sec:3} discusses a similar distortion analysis for the current Phase B design of MAORY~\citep{diolaiti2016maory}, the MCAO module for ELT, and we compare the results with those obtained for GeMS. Final remarks conclude the paper.
\section{GeMS optical design and distortion}
\label{sec:1}
\begin{figure}
	\includegraphics[width=\columnwidth]{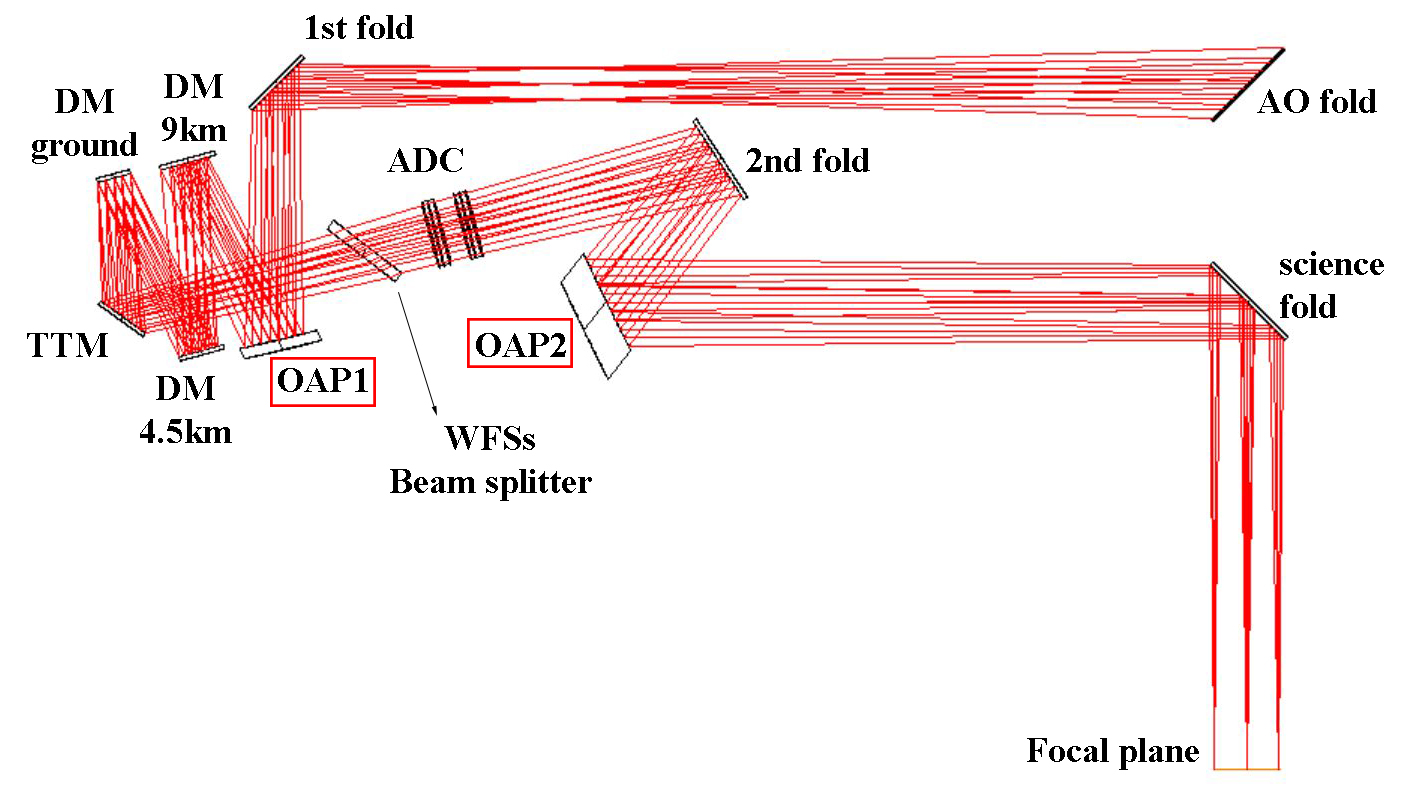}
    \caption{GeMS optical design of the science channel. The three DMs are optically conjugated at the ground and at the atmospheric altitudes of 9km and 4.5km.}
    \label{fig:1}
\end{figure}

\begin{figure}
	\includegraphics[width=\columnwidth]{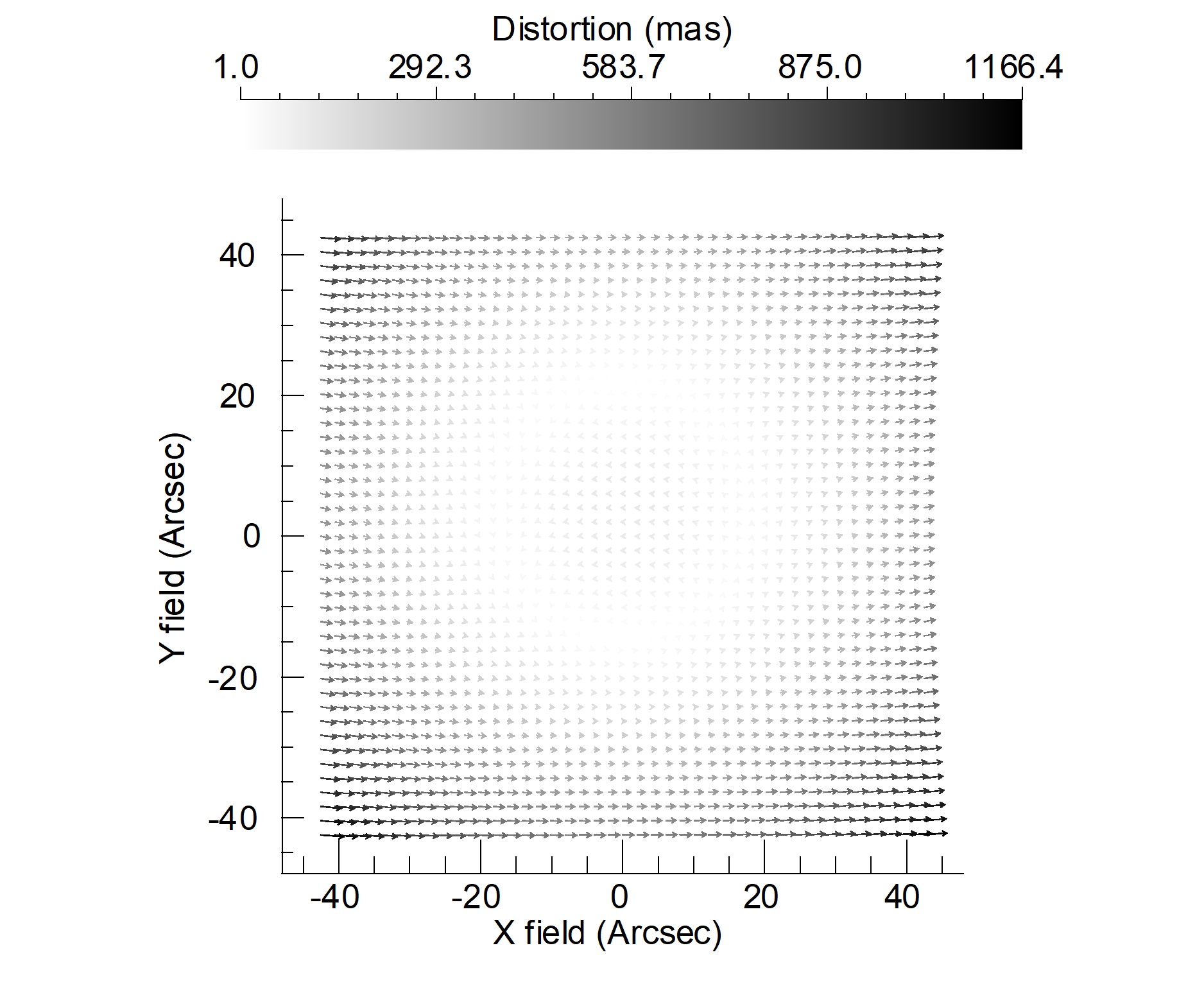}
    \caption{Mapping error related to the GeMS optical design. The GSAOI 85''$\times$85'' FoV is considered. The vectors lengths have been scaled to highlight the distortion structures.}
    \label{fig:2}
\end{figure}

The GeMS is mounted to the Gemini Instrument Support Structure (ISS) co-rotating with the sky at the Cassegrain focus of the Gemini south observatory. Through the ISS, the Gemini telescope F/16 beam is re-directed to the GeMS optical relay via the flat Adaptive Optics (AO) folding mirror. The MCAO wavefront sensing scheme is based on five Laser Guide Stars (LGSs) and three Natural Guide Stars (NGSs). According to this scheme, it is possible to divide the GeMS optical design into three sub-systems:
\begin{enumerate}
\item The science channel which changes the telescope focal length producing a flat, F/33.2 focal plane to the exit port for the science instruments.
\item The NGS channel which is created by the light reflected from a beam-splitter inside the science channel. It is the light used by the NGS wavefront sensors (WFSs).
\item The LGS channel which is created by the light reflected from a beam-splitter inside the NGS channel. It is the light used by the LGS WFSs.
\end{enumerate}
Figure~\ref{fig:1} shows the optical design~\citep{james2003design} of the science channel which has been used to analyse the optical distortion. The design is composed by several optical elements, among which only two with optical power. These are off-axis paraboloid (OAP) mirrors which enclose three deformable mirrors (DMs), a fast tip-tilt mirror (TTM), a beam-splitter that reflects the light used by the WFSs and transmits the near-infrared light, the Atmospheric Dispersion Corrector (ADC) and two folding mirrors used to package the optics. 
The two OAPs, labelled OAP1 and OAP2 in Figure~\ref{fig:1}, form an optical relay where OAP1 creates the image of the telescope secondary mirror (the telescope aperture stop) and OAP2 produces a F/33.2 beam on a flat focal plane. A paraboloid mirror has two focal points, one of which is at infinity. Any chief ray that passes through the focus at infinity, results in an image free of astigmatism and there is no spherical aberration. Adding a second paraboloid (OAP2) mirror to the design, the coma aberration from OAP1 is corrected and the off-axis properties of the two mirrors allow an un-obscured design. To provide a focal plane perpendicular to the central ray (the optical axis), the OAPs are tilted about the telescope focal plane. This design provides a diffraction limited optical quality at the exit port over a relative large Field of View (2 arc-minutes in diameter) but it suffers from mapping errors. Optical distortions are characteristic of unobstructed reflective designs. With only two OAPs, there are few degrees of freedom to play with in order to optimize the design. It is impossible to a have a diffraction limited, perpendicular to the optical axis and distortion-free focal plane. The data used to analyse the optical distortions, have been recorded through a commercial ray-tracing software and user-defined scripts. The atmospheric turbulence is not considered and the DMs are supposed to be rigid mirrors. This means all the considered PSFs are diffraction limited.\\ 
Given a regular grid of $N$ field points on sky $(x_{r},y_{r}\:;\: r=1,...,\:N)$ and their image coordinates at the GeMS focal plane $(x_{d},y_{d})$, the mapping error is defined as:
\begin{equation}
\begin{split}
\delta x=(x_{r}-x_{d})\\ \delta y=(y_{r}-y_{d})
\end{split}
\end{equation}
It is lower then 2\% as shown in Figure~\ref{fig:2} and the amplitude of the distortions in the $x-$component is significantly larger than that in the $y-$component. This is clearly because the optical system is not rotationally symmetric, then the distortion is not radial.\\
In order to model the distortion, it is common to use a least squares fit with high-order polynomial. The independent variables are the reference star coordinates $(x_{r},y_{r})$ while the dependent variables are distorted star coordinates $(x_{d},y_{d})$. The following analyses use the polynomials defined as:
\begin{equation}
\label{eq:2}
\begin{split}
x_{f} = \sum^n_{i=0}\sum^n_{j=0}\:K^{i,j}_x\:x^i_{r}\:y^j_{r}\\
y_{f} = \sum^n_{i=0}\sum^n_{j=0}\:K^{i,j}_y\:x^i_{r}\:y^j_{r}
\end{split}
\end{equation}
The polynomial order $n$ indicates the maximum of the sum of the exponents $i$ and $j$ in the equations. The $K^{i,j}_x$ and $K^{i,j}_y$ are $(n+1)$ square matrices used to model the distortions. The residual distortion map is given by $(x_{r}-x_{f})$, $(y_{r}-y_{f})$ and the astrometric residual error is defined as:
\begin{equation}
\label{eq:3}
RSS_{xy} =  \sqrt{\sigma^2_{x}+\sigma^2_{y}}
\end{equation}
It is the Root Sum of Squares (RSS) $\sigma_{x}$, $\sigma_{y}$ specified as the $x,y$ standard deviations of the residual distortion map. 
\section{Optical distortion analysis}
\label{sec:2}
As formally described in the previous Section, the instrument distortion solution is solved by comparing the positions of detected stars with the corresponding ones in a distortion-free reference catalog. In our analysis, we use a set of reference sources in the so-called object space. The PSF centroids coordinates of these sources are measured in the image space after a sequential ray-tracing from sources through optics to focal plane. In order to match the reference catalog coordinates with the detector coordinates, a linear transformation has to be applied to take into account rigid shifts, rotations and different scale. For this reason, only contributions of second or higher order to the final solution are called distortions. The most common way to solve the GeMS distortion is to consider each GSAOI detector chip independently, obtaining a separate solution for each chip. This approach, when allowed by a sufficient number of detected stars, can enable to better handle potential intra-chip effects. Moreover, the separate solution of individual chip avoids dangerous extrapolations across the chip gaps as it has been first shown by~\cite{anderson2006ground}. In the following analysis, there are no measurement errors and the positions on the detector are known with numerical precision.
\subsection{Static distortion analysis}
\label{sec:2_1}
\begin{figure}
	\includegraphics[width=\columnwidth]{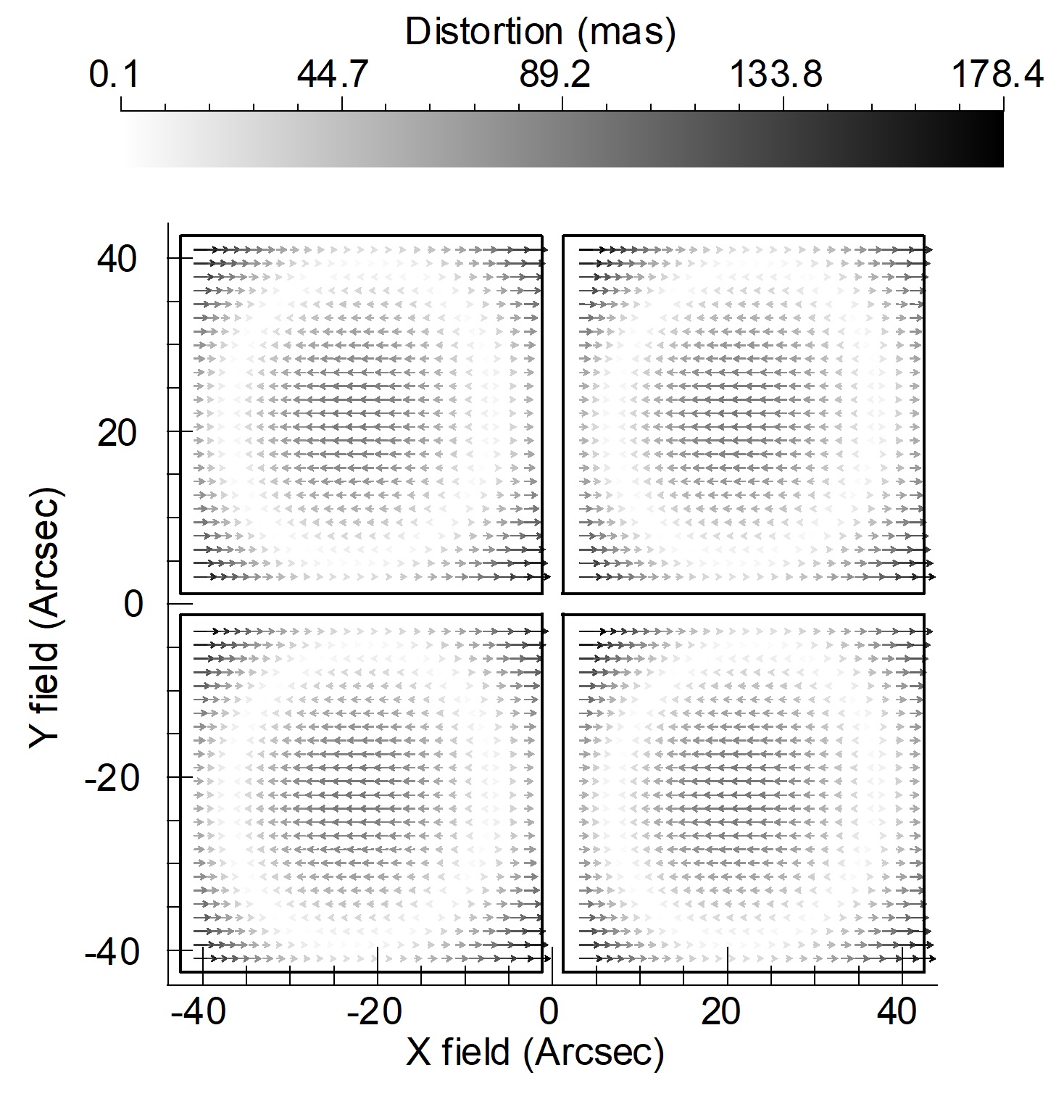}
    \caption{Residual distortion map. The four chips of GSAOI are treated independently. These are contributions of second or higher order to the mapping error shown in Figure~\ref{fig:2}. The vectors lengths have been scaled to highlight the distortion structures.}
    \label{fig:3}
\end{figure}

\begin{figure}
\centering
	\includegraphics[width=\columnwidth]{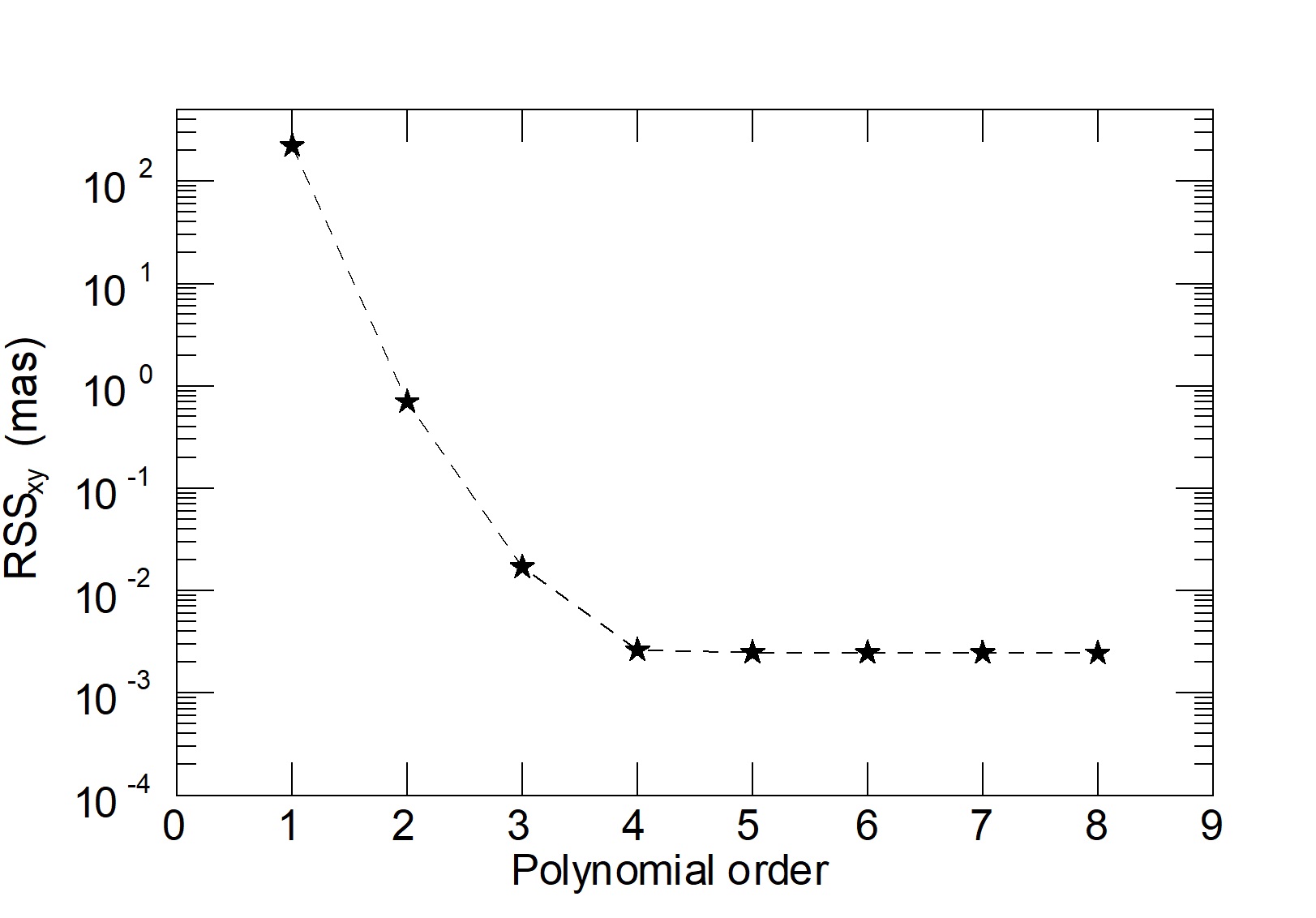}
    \caption{Astrometric residual errors as defined in equation~\ref{eq:3} after fitting a polynomial of order $1\leq n\leq8$.}
    \label{fig:4}
\end{figure}

In the ray-tracing analysis, the GSAOI FoV has been divided into four parts simulating the detector chips. The PSF centroids coordinates were first corrected using a linear transformation ($1^{th}$ order polynomial of equation~\ref{eq:2}) and then, the resulting positions were corrected with the $5^{th}$ order polynomial.
The distortion map of Figure~\ref{fig:3} is the difference between the $5^{th}$ order and $1^{th}$ order corrections. 
The maximum distortion value is about 180 $mas$ which corresponds to 9 pixels of GSAOI detector and the standard deviation over the whole field is about 34 $mas$ (1.7 pixels). This result is quantitatively very similar to the on-sky analysis by~\cite{massari2016high} and easily comparable to the results of the astrometric mask calibration~\citep{riechert2018gems} which were achieved with diffraction limited PSFs.\\
The common circular structure, located at the centre of each chip, is noticeable and it is only due to the static optical distortions of the two OAPs design. It is worth to mention that the same distortion pattern has been found by~\cite{ammons2016precision} in the Shane AO system~\citep{gavel2014shaneao} which is another two OAPs design. The repeated distortion pattern, very similar in magnitude, comes out when treating the detector chips independently. 
\\Astrometry deals with motions of astronomic sources which can be relative to each other or with respect to an absolute coordinate system (or both). Focusing on `absolute astrometry', Figure~\ref{fig:4} shows the residual error, as defined in equation~\ref{eq:3}, when increasing the polynomial order fit. To derive the distortion solution, a grid of 3000 stars over the entire FoV has been considered achieving the level of micro-arcseconds residuals with a fourth order polynomial fit. Higher orders residuals are not dropping off meaning the polynomial coefficients have reached the numerical precision. The fourth order polynomial floor in the residual errors is an important finding, as previous attempts of modelling GeMS geometric distortions adopted up to fifth order polynomials. This means that other sources of distortions (e.g. optical polishing, atmospheric and MCAO terms) can contribute up to that level.
\subsection{Dynamical distortion analysis}
\label{sec:2_2}
Multi-epoch observations are necessary to determine the motions of science objects. However, multi-epoch astrometry suffers from time-variable distortion which, assuming that the diffraction limit is always reached, is mostly related to opto-mechanical and dynamical instabilities of the instrument. 
To evaluate the performance on multi-epoch observations, a Monte Carlo approach has been followed to simulate random gravity-induced flexures of the instrument. To have enough samples and ensure accurate results, 500 Monte Carlo trials have been generated. The GeMS optical bench, attached to the Cassegrain focus, is subjected to flexures when the telescope is tracking. Moreover, the Gemini instruments are mounted and dismounted between runs, modifying the alignment condition. To simulate this variable perturbations, the optical bench has been displaced in the following ranges:
\begin{equation}
\begin{split}
(\Delta x,\:\Delta y) = \pm 5\:mm \\ (\Delta\theta_{x},\:\Delta\theta_{y}) = \pm 0.5\:^\circ
\end{split}
\label{eq:4}
\end{equation}
These are indicative but realistic displacements which degrade the nominal optical quality. The nominal optical design has diffraction-limited performance in each of the J, H and K bands. By looking at the worst Monte Carlo realization in terms of optical quality, it still delivers diffraction-limited PSFs. We followed a conservative approach in the Monte Carlo simulations, since they are based on a parabolic more than a classical Gaussian distribution. This means that the extracted values are more likely at the extreme ends of the chosen range. 
\begin{figure}
	\includegraphics[width=\columnwidth]{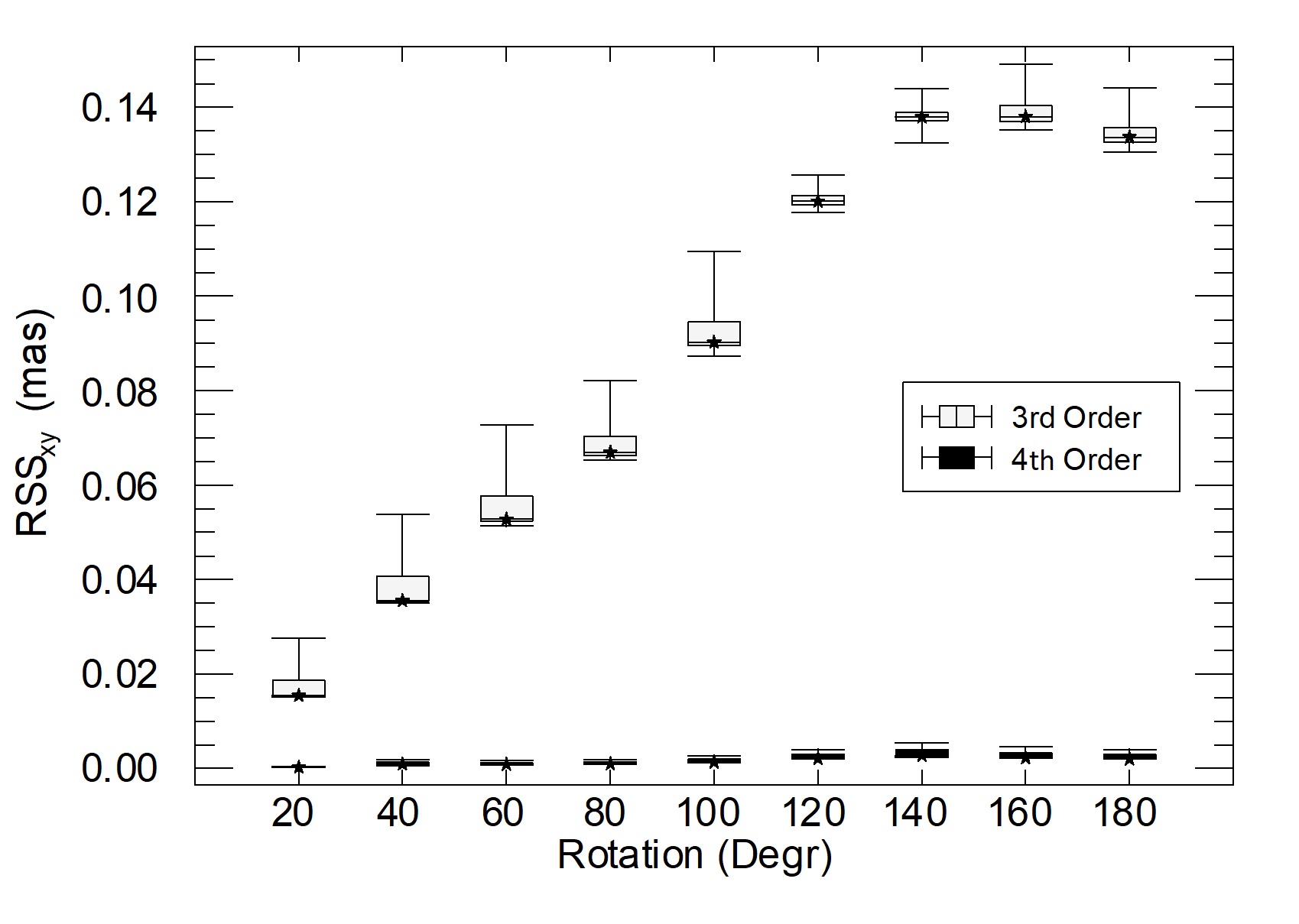}
    \caption{Multi-ephoc astrometric residual errors after the $3^{rd}$ and $4^{th}$ order correction. Box plots show minimum, maximum and quartile values from the Monte Carlo realizations. Each point represents an epoch.}
    \label{fig:5}
\end{figure}
\\For multi-epoch observations the analysis focuses on `relative astrometry' where, given a set of reference sources images at epoch $E$ (which is fixed), the distortion solution has been derived considering the same set of sources at epoch $(E+1)$. Between two successive epochs, a rotation of sky objects of 20$^\circ$ has been considered. Within the GeMS optical design, that means the sources light beams rotate over the optical surfaces. Both, mechanical flexures and beams rotations change the optical path affecting the distortion pattern. Multi-epoch astrometric residual errors are shown in Figure~\ref{fig:5} after the third and fourth order polynomial correction, respectively. Each point is the residual error as defined in equation~\ref{eq:3}, where the reference star coordinates are those at epoch $E$. Box plot, at each point, depicts the $RSS_{x,y}$ values, through their quartiles, from the Monte Carlo realizations. The box width indicates the dispersion of values which is about 5 $\mu as$ with the third order correction. The skewness of box plots is related to the polynomial correction and the considered statistic which follows a parabolic distribution. Given the geometric distortion map, the third order correction has a lower limit residual. From approximately 20$^\circ$ to 80$^\circ$, 50\% of the Monte Carlo trials are corrected up to this lower limit despite the increasing amplitude of distortions due to bench displacements. The combination of values at the extreme ends of the displacement range leads to larger residuals but there are no peculiar pattern or more affected regions of the detector in different epochs. The result of this analysis indicates, at least, the fourth order correction is needed to achieve the best astrometric performance and the instrument flexures introduce distortions up to the third order. The astrometric residual error after the first order polynomial fit reach values up to 2.5 $mas$ with a standard deviation of about 0.4 $mas$, meaning the distortion map of Figure~\ref{fig:3} can change in the range of $\pm$2.5 $mas$ due to gravity-induced flexures of the instrument. This result is again very similar to the on-sky and astrometric mask analyses of the above-cited papers.
\section{Lesson learned for future MCAO instruments}
\label{sec:3}
\begin{figure}
	\includegraphics[width=\columnwidth]{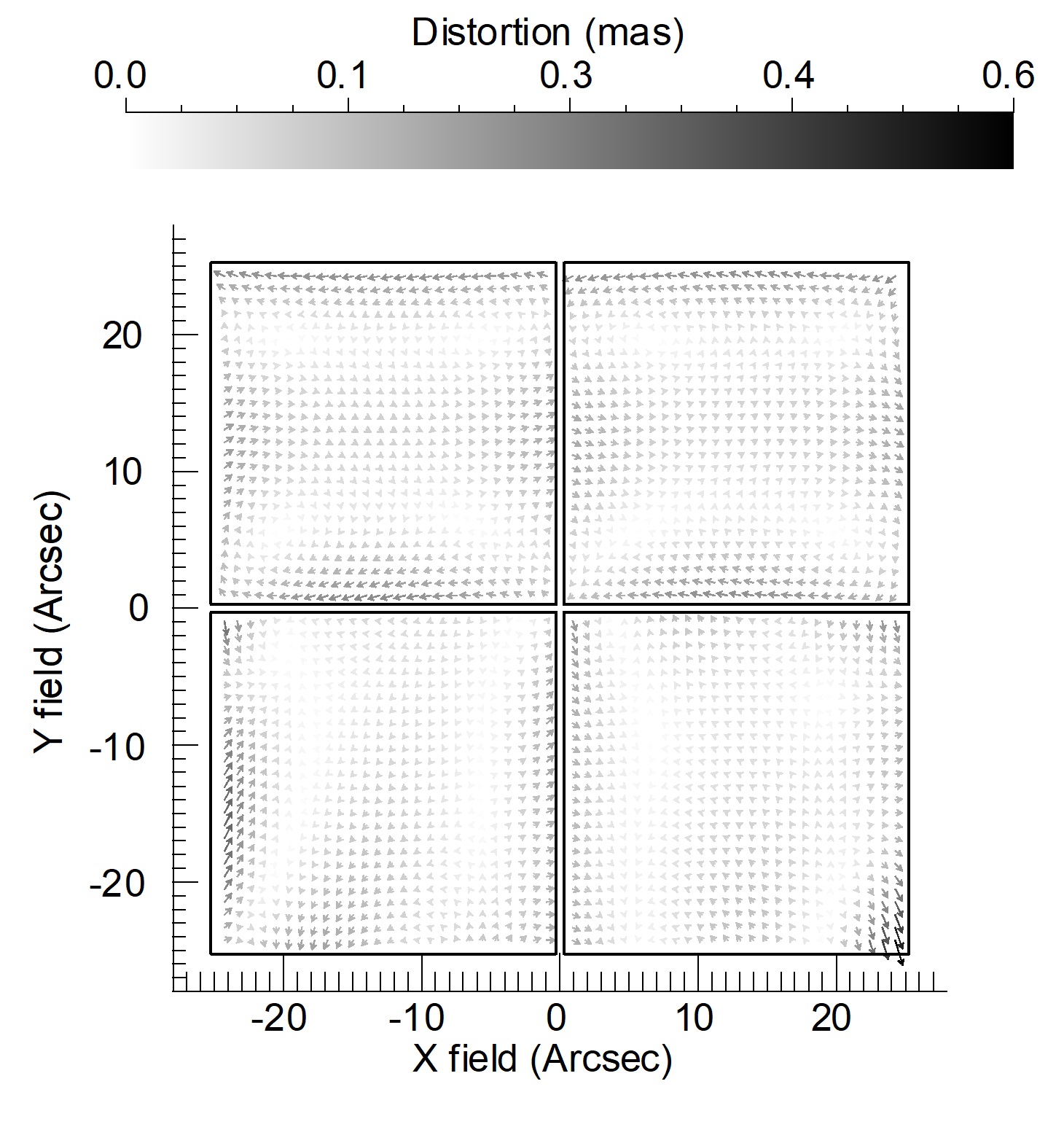}
    \caption{Same of Figure~\ref{fig:3} but considering the 50.5''$\times$50.5'' MICADO FoV at the MAORY exit focal plane.}
    \label{fig:6}
\end{figure}

\begin{figure}
\centering
	\includegraphics[width=0.85\columnwidth]{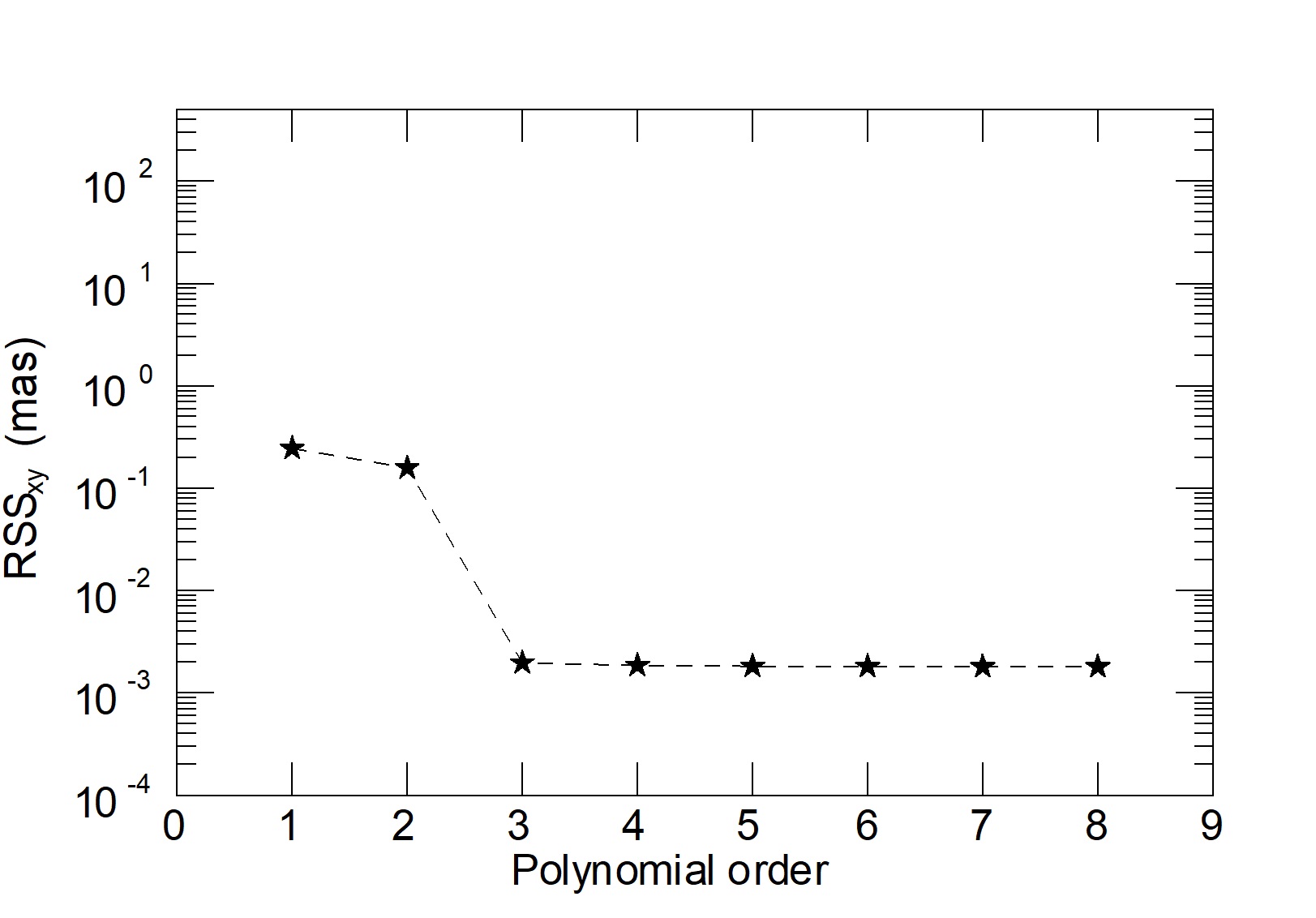}
    \caption{MAORY astrometric residual errors as defined in equation~\ref{eq:3} after fitting a polynomial of order $1\leq n\leq8$.}
    \label{fig:7}
\end{figure}

\begin{figure}
\centering
	\includegraphics[width=0.95\columnwidth]{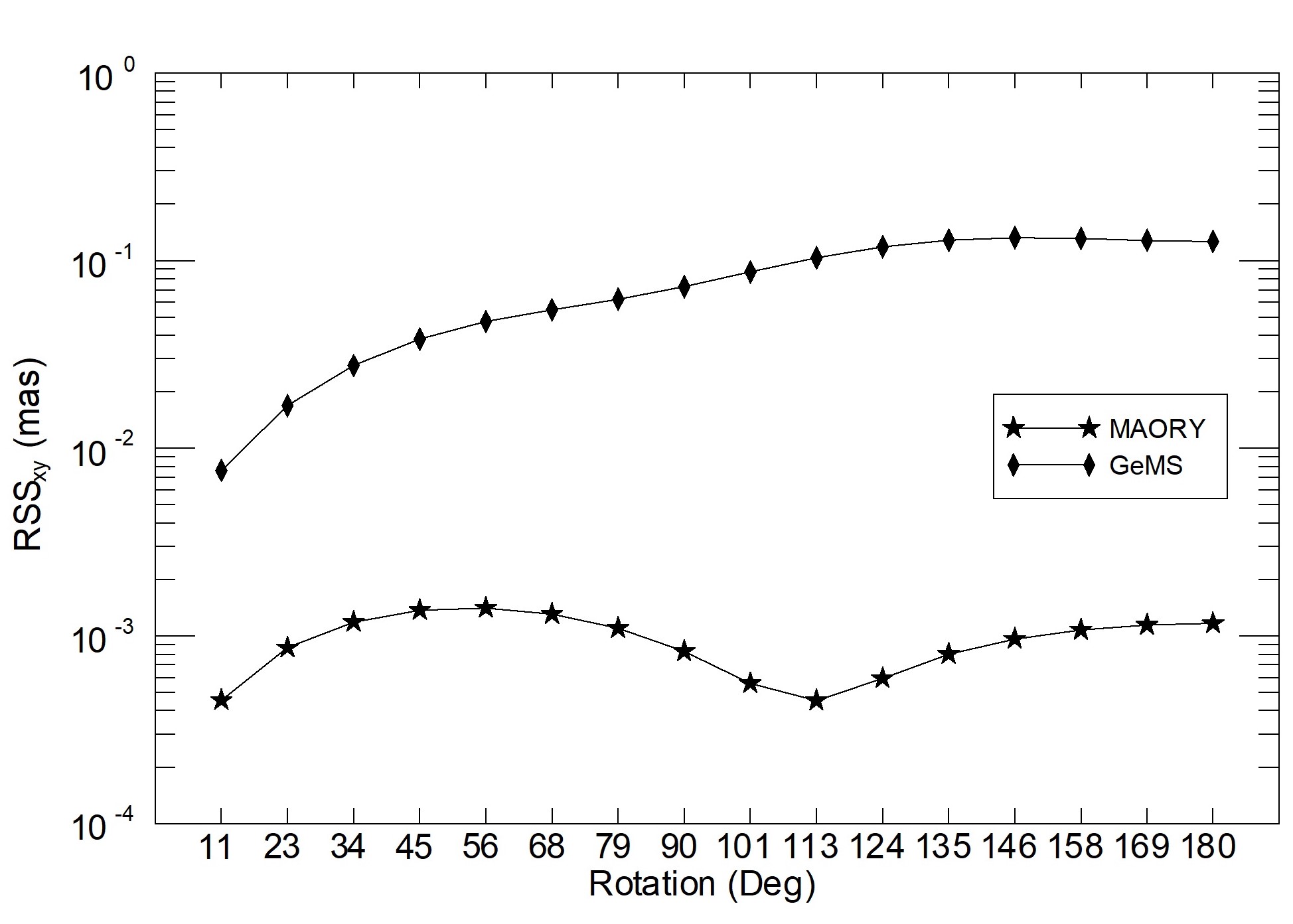}
    \caption{Multi-epoch astrometric residual errors after the $3^{rd}$ order correction for MAORY and GeMS. Each point represents an epoch. The sky rotates from epoch to epoch.}
    \label{fig:8}
\end{figure}

In the next generation of MCAO systems designed for future ELTs, astrometry is one of the main science drivers. 
Thanks to the four to five times smaller diffraction limited PSF with respect to 8 meter class telescopes, and a SNR gain which scales with the third power of telescope diameter, the ELTs' instruments aim for astrometric precision at the level of micro-arcseconds. Optical distortions have to be as low as possible to achieve the high demanding astrometric requirements. This Section briefly reports the case of MAORY that will feed MICADO. MAORY will be placed at the ELT Nasmyth platform and its optics are not co-rotating with the sky during the telescope tracking. In a single-epoch observation, the PSFs at the exit focal plane follows a circular trajectory with small perturbations due to the optical distortions. The sky de-rotation only compensates for the circular part of the trajectory. The residual deviations from a circular trajectory could degrade the PSF, especially if the WFSs detect high order aberrations which are not related to the atmosphere but to the distortion pattern. In this case, the MCAO correction is wrong and a further degradation of the PSF is introduced. The analysis of this effect is beyond the scope of this paper whereas we are interested in a comparative analysis with GeMS in order to make reliable performance predictions on future instruments.\\ 
Figure~\ref{fig:6} shows the residual distortion map at the MAORY exit focal plane over the 50.5''$\times$50.5'' MICADO FoV. These are contributions of second or higher order to the mapping error, equivalent to those of GeMS showed in Figure~\ref{fig:3}. The sub-milliarcsecond level is one of the merit of the current optical design~\citep{lombini2018optical} where the third order polynomial fit is enough to achieve the best astrometric performance as shown in Figure~\ref{fig:7}. The instrument flexures, as in GeMS, are confined to the third order distortion correction while the variation of the distortion pattern is at the level of tens of micro-arcseconds.  Figure~\ref{fig:8} shows the comparison between MAORY and GeMS for multi-epoch astrometric observations. Only the third order correction has been considered to highlight how, unlike GeMS, the third order is sufficient to achieve the required astrometric performance in multi-epoch ''relative astrometry''. A similar approach and result has been achieved by~\cite{rodeghiero2018impact} studying the ELT distortions originated from the optics positioning errors and telescope instabilities.
%
\section{Final remarks}
This work has proved it is possible to simulate, at the design stage, the behaviour of operating instruments. The notable example which is analysed here is the case of GeMS/GSAOI whose astrometric performances have been analysed in detail through on-sky data during the last years. The results, here presented, unveiled the nature of the distortions which are mainly due to the optical design with two OAPs and opto-mechanical instabilities of the instrument. We have assumed that diffraction limit is reached overall the FoV. Thus the PSF shape and variation across the FoV were not taken into account to avoid an additional level of complexity of the analysis that has been focused on the theoretical limit of GeMS astrometric precision. The results are comparable and quantitatively very similar to the outcomes of the astrometric mask calibration highlighting how, an accurate design model is sufficient to validate the astrometric performance of future instruments.\\ The third order distortion solution has demonstrated to be able to correct for the instrument flexures but it is not enough to achieve the best astrometric performances on multi-epoch observations where, at least, a fourth order correction is needed. The result is not only consistent with that coming from the use of the calibration mask, but also with distortion corrections performed directly on the sky. This means that the centroiding measurements of previous works were not affected by systematic errors and affected by negligible intrinsic errors with respect to the optical distortions.\\
The future generation of MCAO instruments take advantage of our analysis and will be designed to significantly reduce the amount of optical distortions which is a significant contributor of a more refined astrometric error budget. Moreover, an accurate design model could be sufficient to develop calibration algorithms in order to interpolate the observational conditions. This approach can avoid any additional hardware on the instruments reducing the engineering/calibration time during operations.
\section*{Acknowledgements}
The authors would like to thank Gaetano Sivo who kindly provided the GeMS optical design used for the ray-tracing analyses and all the GeMS team members for constructive discussions over the last years. We are grateful to the anonymous referee for her/his review work oriented to improve this paper. GF has been supported by the Futuro in Ricerca 2013 (grant RBFR13J716). This work focus on one of the instruments of the Gemini Observatory, which is operated by the Association
of Universities for Research in Astronomy, Inc., under a cooperative agreement with the NSF on behalf of the
Gemini partnership: the National Science Foundation (United States), the National Research Council (Canada),
CONICYT (Chile), Ministerio de Ciencia, Tecnolog\'ia e Innovaci\'on Productiva (Argentina), and Minist\'erio da
Ci\^encia, Tecnologia e Inova\c{c}\~ao (Brazil).




\bibliographystyle{mnras}
\bibliography{Biblio} 

\begin{thebibliography}{}
\makeatletter
\relax
\def\mn@urlcharsother{\let\do\@makeother \do\$\do\&\do\#\do\^\do\_\do\%\do\~}
\def\mn@doi{\begingroup\mn@urlcharsother \@ifnextchar [ {\mn@doi@}
  {\mn@doi@[]}}
\def\mn@doi@[#1]#2{\def\@tempa{#1}\ifx\@tempa\@empty \href
  {http://dx.doi.org/#2} {doi:#2}\else \href {http://dx.doi.org/#2} {#1}\fi
  \endgroup}
\def\mn@eprint#1#2{\mn@eprint@#1:#2::\@nil}
\def\mn@eprint@arXiv#1{\href {http://arxiv.org/abs/#1} {{\tt arXiv:#1}}}
\def\mn@eprint@dblp#1{\href {http://dblp.uni-trier.de/rec/bibtex/#1.xml}
  {dblp:#1}}
\def\mn@eprint@#1:#2:#3:#4\@nil{\def\@tempa {#1}\def\@tempb {#2}\def\@tempc
  {#3}\ifx \@tempc \@empty \let \@tempc \@tempb \let \@tempb \@tempa \fi \ifx
  \@tempb \@empty \def\@tempb {arXiv}\fi \@ifundefined
  {mn@eprint@\@tempb}{\@tempb:\@tempc}{\expandafter \expandafter \csname
  mn@eprint@\@tempb\endcsname \expandafter{\@tempc}}}

\bibitem[\protect\citeauthoryear{Ammons et~al.,}{Ammons
  et~al.}{2016}]{ammons2016precision}
Ammons S.~M.,  et~al., 2016, in Adaptive Optics Systems V. p. 99095T

\bibitem[\protect\citeauthoryear{Anderson, Bedin, Piotto, Yadav  \&
  Bellini}{Anderson et~al.}{2006}]{anderson2006ground}
Anderson J.,  Bedin L.~R.,  Piotto G.,  Yadav R.~S.,   Bellini A.,  2006,
  Astronomy \& Astrophysics, 454, 1029

\bibitem[\protect\citeauthoryear{{Beckers}}{{Beckers}}{1988}]{beckers88}
{Beckers} J.~M.,  1988, in ESO Conference on Very Large Telescopes and their
  Instrumentation. pp 693--703

\bibitem[\protect\citeauthoryear{{Beckers}}{{Beckers}}{1989}]{beckers89a}
{Beckers} J.~M.,  1989, in Active Telescope Systems. pp 215--217

\bibitem[\protect\citeauthoryear{Bernard, Neichel, Mugnier  \& Fusco}{Bernard
  et~al.}{2018}]{bernard2018optimal}
Bernard A.,  Neichel B.,  Mugnier L.,   Fusco T.,  2018, Monthly Notices of the
  Royal Astronomical Society, 473, 2590

\bibitem[\protect\citeauthoryear{Carrasco et~al.,}{Carrasco
  et~al.}{2012}]{carrasco2012results}
Carrasco E.~R.,  et~al., 2012, in Adaptive Optics Systems III. p. 84470N

\bibitem[\protect\citeauthoryear{Dalessandro et~al.,}{Dalessandro
  et~al.}{2016}]{dalessandro2016gems}
Dalessandro E.,  et~al., 2016, The Astrophysical Journal, 833, 111

\bibitem[\protect\citeauthoryear{Davies et~al.,}{Davies
  et~al.}{2018}]{davies2018micado}
Davies R.,  et~al., 2018, in Ground-based and Airborne Instrumentation for
  Astronomy VII. p. 107021S

\bibitem[\protect\citeauthoryear{Diolaiti et~al.,}{Diolaiti
  et~al.}{2016}]{diolaiti2016maory}
Diolaiti E.,  et~al., 2016, in Adaptive Optics Systems V. p. 99092D

\bibitem[\protect\citeauthoryear{{Fiorentino} et~al.,}{{Fiorentino}
  et~al.}{2017}]{fiorentino2017}
{Fiorentino} G.,  et~al., 2017, arXiv e-prints, \href
  {http://adsabs.harvard.edu/abs/2017arXiv171204222F} {}

\bibitem[\protect\citeauthoryear{{Gaia Collaboration} et~al.,}{{Gaia
  Collaboration} et~al.}{2016}]{refId0}
{Gaia Collaboration} et~al., 2016, \mn@doi [A\&A]
  {10.1051/0004-6361/201629272}, 595, A1

\bibitem[\protect\citeauthoryear{Gavel et~al.,}{Gavel
  et~al.}{2014}]{gavel2014shaneao}
Gavel D.,  et~al., 2014, in Adaptive Optics Systems IV. p. 914805

\bibitem[\protect\citeauthoryear{{Gillessen} et~al.,}{{Gillessen}
  et~al.}{2017}]{gillessen17}
{Gillessen} S.,  et~al., 2017, \mn@doi [\apj] {10.3847/1538-4357/aa5c41}, \href
  {http://adsabs.harvard.edu/abs/2017ApJ...837...30G} {837, 30}

\bibitem[\protect\citeauthoryear{James, Boyer, Buchroeder, Ellerbroek  \&
  Hunten}{James et~al.}{2003}]{james2003design}
James E.,  Boyer C.,  Buchroeder R.~A.,  Ellerbroek B.~L.,   Hunten M.~R.,
  2003, in Adaptive Optical System Technologies II. pp 67--81

\bibitem[\protect\citeauthoryear{Lombini et~al.,}{Lombini
  et~al.}{2018}]{lombini2018optical}
Lombini M.,  et~al., 2018, in Optical Design and Engineering VII. p. 1069011

\bibitem[\protect\citeauthoryear{Massari et~al.,}{Massari
  et~al.}{2016a}]{massari2016astrometry}
Massari D.,  et~al., 2016a, Astronomy \& Astrophysics, 595, L2

\bibitem[\protect\citeauthoryear{Massari et~al.,}{Massari
  et~al.}{2016b}]{massari2016high}
Massari D.,  et~al., 2016b, in Adaptive Optics Systems V. p. 99091G

\bibitem[\protect\citeauthoryear{Neichel, Lu, Rigaut, Ammons, Carrasco  \&
  Lassalle}{Neichel et~al.}{2014}]{neichel2014astrometric}
Neichel B.,  Lu J.~R.,  Rigaut F.,  Ammons S.~M.,  Carrasco E.~R.,   Lassalle
  E.,  2014, Monthly Notices of the Royal Astronomical Society, 445, 500

\bibitem[\protect\citeauthoryear{Riechert, Garrel, Pott, Sivo  \&
  Marin}{Riechert et~al.}{2018}]{riechert2018gems}
Riechert H.,  Garrel V.,  Pott J.-U.,  Sivo G.,   Marin E.,  2018, in
  Ground-based and Airborne Instrumentation for Astronomy VII. p. 1070232

\bibitem[\protect\citeauthoryear{Rigaut et~al.,}{Rigaut
  et~al.}{2013}]{rigaut2013gemini}
Rigaut F.,  et~al., 2013, Monthly Notices of the Royal Astronomical Society,
  437, 2361

\bibitem[\protect\citeauthoryear{Rodeghiero, Pott, Arcidiacono, Massari,
  Gl{\"u}ck, Riechert  \& Gendron}{Rodeghiero
  et~al.}{2018}]{rodeghiero2018impact}
Rodeghiero G.,  Pott J.-U.,  Arcidiacono C.,  Massari D.,  Gl{\"u}ck M.,
  Riechert H.,   Gendron E.,  2018, Monthly Notices of the Royal Astronomical
  Society

\bibitem[\protect\citeauthoryear{Trippe, Davies, Eisenhauer, Schreiber, Fritz
  \& Genzel}{Trippe et~al.}{2010}]{trippe2010high}
Trippe S.,  Davies R.,  Eisenhauer F.,  Schreiber N.~F.,  Fritz T.,   Genzel
  R.,  2010, Monthly Notices of the Royal Astronomical Society, 402, 1126

\bibitem[\protect\citeauthoryear{Yelda, Lu, Ghez, Clarkson, Anderson, Do  \&
  Matthews}{Yelda et~al.}{2010}]{yelda2010improving}
Yelda S.,  Lu J.~R.,  Ghez A.~M.,  Clarkson W.,  Anderson J.,  Do T.,
  Matthews K.,  2010, The Astrophysical Journal, 725, 331

\makeatother
\end{thebibliography}


\bsp	
\label{lastpage}
\end{document}